# Reliable generation of isomorphic physics problems using Generative AI with prompt-chaining and tool use

Zhongzhou Chen
*Department of Physics, University of Central Florida, 4111 Libra Drive, Orlando, Florida, USA 32816*

We present a method for generating large numbers of isomorphic physics problems using generative AI services such as ChatGPT, through prompt chaining and tool use. This approach enables precise control over structural variations—such as numeric values and spatial relations-while supporting diverse contextual variations in the problem body. By utilizing the Python code interpreter, the method supports automatic solution validation and simple diagram generation, addressing key limitations in existing LLM-based methods. We generated two example isomorphic problem banks and compared the outcome against two simpler prompt-based approaches. Results show that prompt-chaining produces significantly higher quality and more consistent outputs than simpler, non-chaining prompts. We also show that GenAI services can be used to validate the quality of the generated isomorphic problems. This work demonstrates a promising method for efficient and scalable problem creation accessible to the average instructor, which opens new possibilities for personalized adaptive testing and automated content development.

## I. INTRODUCTION

There has been significant progress in developing Automated Question Generation (AQG) and Automated Item Generation (AIG) technologies in education over the past decade. These technologies aim to reduce the time and cost of item creation while increasing the availability of questions for both assessment and practice [1].

Early AQG/AIG approaches primarily relied on hard-coded, template-based methods, which were often time-consuming to develop and required domain-specific programming [2]. More recent research has shifted toward leveraging large language models (LLMs). For example, Dijkstra et al. trained a GPT-3 model to generate multiple-choice questions with valid answers and plausible distractors [11]. Jiao et al. proposed an energy-based model to generate math word problems with appropriate difficulty levels [12]. Chan et al. prompted GPT-3.5 and GPT-4 to produce questions in chemistry, physics, and mathematics [3]. Omopekunola and Kardanova used simple prompts to generate physics multiple-choice items aligned with Bloom's taxonomy levels [4].

Most of those studies generated new questions within a given content area and cognitive level in Bloom's taxonomy. However, as pointed out in Maity et al. [5] , LLMs often struggle to control difficulty and cognitive complexity, frequently defaulting to factual recall questions when operating under such general constraints. In addition, LLMs could sometimes hallucinate on the correct answer, especially for numeric calculation problems.

An alternative approach to AQG is the creation of **isomorphic problem variants**—questions that assess the same underlying concepts and principles but differ in surface features. This was traditionally achieved through template-based generation and applying structured constraints to control variation. For instance, Arendasy and Sommer [6] generated algebra word problems by combining required and optional sentences and manipulating the number of equations and unknowns to maintain construct equivalence. Isomorphic problems offer better control over item difficulty and reduce construct-irrelevant variance, making them particularly useful in contexts such as repeated assessments [7] and deliberate practice activities [8,9].

Despite these advantages, few studies have explored the use of LLMs for generating isomorphic problems with precise control over certain variations such as changing the problem context or the details of the solution process. One challenge is that LLMs can struggle to consistently follow instructions on certain variations while maintaining creativity in other aspects of the problem. One example in this direction is Norberg et al. [10], who used GPT-4 to rewrite explanations for math word problems in pre-defined styles, relying on the model's code-generation capabilities to ensure solution correctness via Python. A second challenge is the generation of diagrams for isomorphic problem sets, since current LLMs have limited capabilities in precisely controlling visual elements, making diagram generation a persistent obstacle.

In this paper, we introduce a reliable method for generating and validating large numbers of isomorphic problems using Generative AI services such as ChatGPT, through prompt chaining and tool-use, which achieves precise control over variations and contextual creativity simultaneously.

Prompt chaining is a prompt engineering technique in which a complex task is broken down into multiple sub-tasks, and executed through a chain of prompts [11]. Outputs from earlier prompts are used as context for subsequent prompts, enabling complex, multi-step reasoning tasks that surpass the limitations of single-prompt interactions. In isomorphic problem generation, prompt chaining is used to separate construct relevant variations from construct irrelevant variations, allowing for different types of constraints to be applied independently.

Tool-use refers to the ability of a language model to interact with external tools or functions to perform tasks beyond natural language generation. In particular, the Python code interpreter [12] plugin allows the model to write and execute Python scripts in real time in response to a natural language prompt. The python interpreter tool is either used automatically by ChatGPT, or being explicitly requested in the prompt. In the context of isomorphic problem generation, this tool can be used to systematically generate variations, evaluate answer correctness, and produce diagrams that align precisely with the problem statement.

We introduce a seven-step process for LLM assisted isomorphic problem generation using Generative AI in the Methodology section, using two examples of creating isomorphic problem banks: one for numerical computation questions and one for conceptual multiple choice questions with diagrams. We will also compare the output with single prompt generation to demonstrate the strength of prompt-chaining, and discuss future development possibilities in the conclusions section.

## II. METHODS

### A. LLM assisted isomorphic problem generation

The generation process consists of seven steps:

Step 1: Identify a template problem or template problem type.

Step 2: Identify the components of the problem.

Step 3: Define *structural variations* and *contextual variations* (defined below), and determine the constraints for each.

Step 4: Design a prompt chain to generate structural and contextual variations for individual components.

Step 5: Execute the prompt-chain, and iteratively improve each prompt based on the outcome.

Step 6: Combine individual components into complete problems, and output in desired format.

Step 7: Verify the correctness of the generation outcomes using Generative AI.

Here we use **structural variations** to refer to construct-relevant variations to the core structure of the problem that must stay within a more precise, user-defined range to ensure the correctness and/or the appropriate difficulty of the problem. Common examples include numerical values of variables, spatial arrangement of objects, number of objects (forces, particles, etc.) in the problem.

Structural variations are often created by a combination of direct LLM generation and use of the python interpreter tool. For example, in generating numerical values for variables, the python interpreter can be used to pose strict constraints on the numerical values. For example, ChatGPT can generate python code to ensure the total applied horizontal force on an object is less than or greater than the maximum static friction between the object and the surface.

Contextual variations refer to variations to the surface features of a problem. Those variations are subjected to less stringent restrictions yet require creativity from the LLM. Constraints to the contextual variations could take into account additional factors aside from the correctness of the problems, such as the student population's reading level, language proficiency, and cultural background/life experience. The most common contextual variation is the context of a problem.

Structural variations and contextual variations can interact with each other. For example, the numerical value of the weight of the object (structural variation) should be constrained by the type of object in the problem (contextual variation). Those interactions are a key consideration in the design of the prompt chain.

### B. Prompt-chain design of isomorphic problem bank generation

Below we detail the problem components, structural and contextual variations, and the design of prompt-chain for two example problem banks for this study.

Since some of the prompts are quite lengthy, we only explain the goal of each prompt here and include snippets of example prompt text. The actual prompts used to create the problem banks can be found in Appendix I.

<u>*Problem bank 1:*</u> contains a numerical calculation problem with no corresponding diagram.

All problems in the bank follows the same template: An object is being pushed or pulled by a single force at an angle on a rough surface at constant velocity. Each problem asked students to calculate the numerical value of either mass, force, or coefficient of kinetic friction.

Problem Components: Problem body, known variables, unknown variable, Problem solution.

Contextual variation and constraints: Common scenarios involving pushing or pulling of an object on rough surface with an angled force.

Structural variations:

1. The direction and nature of the external force (pointing up or down, pushing or pulling). 2. The values of variables, including coefficient of friction, angle of force, mass of object, and magnitude of force. 3. The selection of unknown variables.

Structural variation constraints:

1. The angle of the force varies between 10 and 60 degrees in both positive and negative direction with respect to the horizontal direction. (This constraint is to avoid extreme and unphysical values. The exact boundaries are arbitrary) 2. The mass and force should be appropriate for objects in the problem context. 3. The horizontal component of external force should balance the kinetic friction. 4. The magnitude of force must be positive.

The unknown variable should be selected from the force, mass, or the coefficient of kinetic friction, but not the angle of the force so as to maintain relatively similar mathematical complexity for the solution.

<u>*Prompt-chain design:*</u> The prompt chain consists of five prompts. For this case, contextual variation is generated before structural variation, since the context informs the choice of variable values.

Prompt 1: asks GenAI to generate 10 different problem contexts suitable for the current problem type, following a few examples. The context examples were written briefly such as:

```
horse pulling a sledge on snow. upwards.
A person pushing a couch across carpeted floor. Downwards.
```

The last word indicates the direction of the pulling or pushing force in the context.

Prompt 2: asks GenAI to generate random values of all variables in each problem, adhering to the structural constraints listed above. The variables generated should be appropriate to the previously generated context. Then the prompt asks AI to calculate the required force for constant velocity and check if it is within the estimated force range appropriate for the context. We show the last part of the prompt which ChatGPT consistently utilizes python to execute:

```
4. In addition, generate a generous estimation of the force that can be exerted by the human, animal or machine doing the pushing or pulling (not the object).
 Next, calculate the magnitude of the force according to:

 F = \mu * m *g/(cos(\theta) + \mu*sin(\theta)).

 Check if the magnitude of F (in Newtons) is within the estimated range, and if the magnitude is positive. If yes, list the values. If not, re-generate the random numbers.
```

Prompt 3: Select a different unknown variable for each problem, and write the problem body according to the results of previous steps. This prompt is iterated multiple times, and specific requirements were added until the output was satisfactory. For example, two of the specific requirements are as follows:

```
    6. Ask students to calculate the unknown variable, and specify the unit. For example, find the force F in units of Newtons.
    7. Specify significant figures required for the answer.
```

Prompt 4: Write the problem solution according to a list of user-specified instructions. This prompt is developed in similar fashion as prompt 3, with multiple iterations.

Prompt 5: Output both the problem body and solution into a pre-defined data format.

<u>*Problem bank 2*</u> consists of conceptual physics problems involving a diagram.

Template Problem Type: The problems ask students to compare three parabolic trajectories of projectile motion, with the same starting position but different height and range. Students are asked to choose the correct relation between the flight times of the three projectile motion trajectory (see the Results section and Figure 1 for example problems). The correct answer is that the flight time relation corresponds to the relation between the heights of the three trajectories.

Problem components: Problem body, diagram, correct choice item, distractors.

Contextual variation: Scenarios involving three projectiles. The objects should not be subjected to significant air resistance such as balloons or arrows.

Structural variations: 1. The correct answer can be any possible relations between three items. 2. No two trajectories should have the same height and same range, to avoid visual overlapping. 3. The range relation should be different from the height relation. 4. One of the distractors should reflect the relation between the ranges of the trajectories. 5. The three ranges and the three heights shouldn't be all identical. 6. Each problem should have four choice items. 6. None of the choice items should have non-determinant relational forms such as "A < B > C", or "B > A < C" 7. For the problem diagram, the trajectories should be sufficiently different visually to avoid overlap, but not too different for all trajectories to be clearly visible.

The prompt chain in this case involves more prompts to handle more complicated structural variation that affects both the figure and the distractors. The prompt chain also utilized a data table as a form of storing and transferring information between different prompts in the prompt chain.

Prompt 1: Generate all possible relations between three trajectories, A, B and C, to be used as flight time relation choice item. The core part of the prompt is as follows:

```
First, please list all possible relations between the flight time of the three trajectories, for example, A > B > C, or A = B < C.
```

Prompt 2: For each flight time relation, generate 2 range relations by replacing both relational symbols. Store all relations in a table format. (For example, change A > B > C into A < B = C).

Prompt 3: Add a new columns to the previously generated table, containing random numbers for trajectory height and trajectory range. The numbers should correspond to the relations in each row, and reflect structural constraint 7.

Prompt 4: Generate downloadable diagrams using python and matplotlib according to the parameters stored in each row of the previous table. A snipped of the prompt is shown here:

```
Now write a program using python and matplotlib to generate one diagram for each row in the above table, according to the parameters in the Table. In each diagram, the three trajectories will be represented by parabolic lines using dashed lines.
```

Prompt 5: For each problem, generate two more random distractors that are different from the flight time relation and the range relation, and avoid non-determinant forms. The main part of the prompt is as follows:

```
add two columns named "Distractor I" and "Distractor II". Each column should contain one random relation between A, B and C that is different from either the flight time or the range. Avoid non-determinant forms such as "A < B > C", or "B > A < C"
```

Prompt 6: Write 10 possible scenarios involving three projectiles:

```
Next, come up with 10 possible situations that involve projectile motion of three objects, all three projectiles should start from ground level and land on ground level. For example, three boys throwing rocks, three gunships firing guns, etc. Avoid objects that might be subjected to air resistance, such as arrows.
```

Prompt 7: Randomly associate the 10 possible scenarios to the entire set of choice items, by adding a new column in the data table.

Prompt 8: Write the problem bodies of the 26 problems.

Prompt 9: Write the problem bodies and choice items and store them in the pre-defined data format.

In prompts 2 and 3, the AI is instructed to store outputs in a table format, which increases the reliability of passing information to the next prompt. For prompt 2, each height relation could correspond to more than 2 valid range relations. We chose 2 to reduce the complexity of the step for more accurate outcome. Also, the selection of 10 scenarios in prompt 6 was arbitrary and can be any reasonable number. Empirically, the chances of ChatGPT generating unrealistic or unphysical scenarios increases with the number of different scenarios.

We did not include a prompt to generate the solution text for this problem bank, and instead used the same solution text for all problems.

The data storage format selected for storing the isomorphic problem data is yaml [13]. The specific yaml format is a modification of Yaml Based Exams by Robbert Harms [14]. Using a custom program developed by University of Central Florida Center for Distributed Learning, the modified yaml can be translated into QTI format and imported as item bank into the CANVAS learning management system.

### D. Verifying generation quality using Generative AI

The last step in the process is executed after the problem bank is being generated and all items recorded in the yaml format. This step simply involves uploading the yaml file to a generative AI service (in this case Gemini Pro 2.5), and prompting GenAI to check the correctness of all the answers. For problem bank 1, GenAI was prompted to check the correctness of the final answer. For problem bank 2, GenAI was specifically prompted to check if any two of the answer choices were equivalent to each other. The prompt to check Problem Bank 1 listed here:

```
I have the attached yaml file that contains 20 questions, the correct answer, and the solution. Some of the solution and answers might be incorrect. Can you please check the first five problems, and see if the answer/solutions are correct? If incorrect, state what is incorrect and provide the correct yaml text for that question.
```

### C. One-prompt and Simple prompt testing cases

The one-prompt condition was constructed by combining prompts in the original prompt chain and making minimal edits. This approach ensures that the one prompt used contains the exact same amount of information as the prompt chain. For bank 2, the original prompt chain was combined into two prompts with the first prompt generating downloadable figures using python, and the second prompt generating the problem bodies.

The Simple prompt testing case was created for bank 1, in which we used a simplified prompt to ask chatGPT to generate 10 isomorphic versions of the problem and solution based on one sample problem and a small number of requirements. The sample problem was picked from the prompt-chain generated bank 1. This prompt is designed to mimic the naïve approach that an inexperienced instructor might try for the task. The simple prompt was not conducted for Problem Bank 2, since it is impossible to ask chatGPT to generate the downloadable figures and accurate meaningful distractors using a simple prompt following a single example, therefore we did not perform this test case.

### D. Implementation and Data Availability

Both prompt chains were executed in the ChatGPT interface on the OpenAI website. The verification step (step 7) was executed using Gemini Pro 2.5. All prompts, conversation history, generated problem banks, and comparison tests mentioned in this paper can be access at: https://github.com/Zhongzhou/LLM-Isomorphic-problem-creation.

## III. RESULTS

### A. Problem Bank Generation

Problem Bank 1: All prompts in the prompt chain were responded satisfactorily with the GPT-4o model, with the python code interpreter being used for prompt 2. Prompt 3 was iterated several times to add multiple requirements to ensure the quality of the output. For example, one requirement is: "Imply that the object is moving/sliding at constant or uniform speed."

We initially generated 10 isomorphic variations for the purpose of this study, although there's no upper limit of isomorphic problems that can be generated. One undergraduate research assistant later used the same prompt chain and generated 10 additional isomorphic variations, using Gemini Pro 2.5.

One example problem generated by the LLM is shown here:

```
A worker is pushing a wooden crate across a concrete floor at a uniform speed.
The coefficient of kinetic friction between the crate and the floor is $0.36$.
The worker applies a force at an angle of $12.74^\circ$ downward with respect to the horizontal.
The required force exerted is $465.64$ N. The acceleration due to gravity is $9.81$ m/s².
Find the mass of the crate in kilograms. Express your answer with two significant figures.
answer: 117.67
```

Each problem in the bank has a unique cover story, with context appropriate random numbers, and a correct answer generated by python code. Each problem also has a solution (see Appendix II).

Problem Bank 2: GPT systematically generated all 13 possible relations between three elements, and added two major distractors to each relation, resulting in a problem bank with 26 variations. Prompts 2 and 3 required o3-mini-high model for correct execution, while all other prompts were executed using GPT-4o. Prompt 4 was completed using the python interpreter to generate the diagrams.

Two sample problems generated by the prompt-chain are shown below, with the corresponding diagrams shown in Figure 1. The correct answers are shown in bold font:

```
q-1: Three engineering students test launchers that fire identical steel spheres from the same ground-level point. The three projectiles (A, B, and
```

C) follow distinct parabolic paths and land back on the ground. Their trajectories are shown in the figure.

Which of the following represents the correct ranking of their time of flight?

C < A < B;
**A > B > C;**
A < B < C;
A = C < B;

q-9: During a robotics competition, three autonomous soccer robots kick soccer balls (A, B, and C) into the air from the same spot. The trajectories of the balls are shown in the figure, and each ball returns to the ground. Which of the following correctly ranks the time each ball is in the air?

A < C = B;
B < A = C;
**C > A > B;**
C < A < B

In each problem, one of the distractors corresponds to the relation of the range of the trajectories.

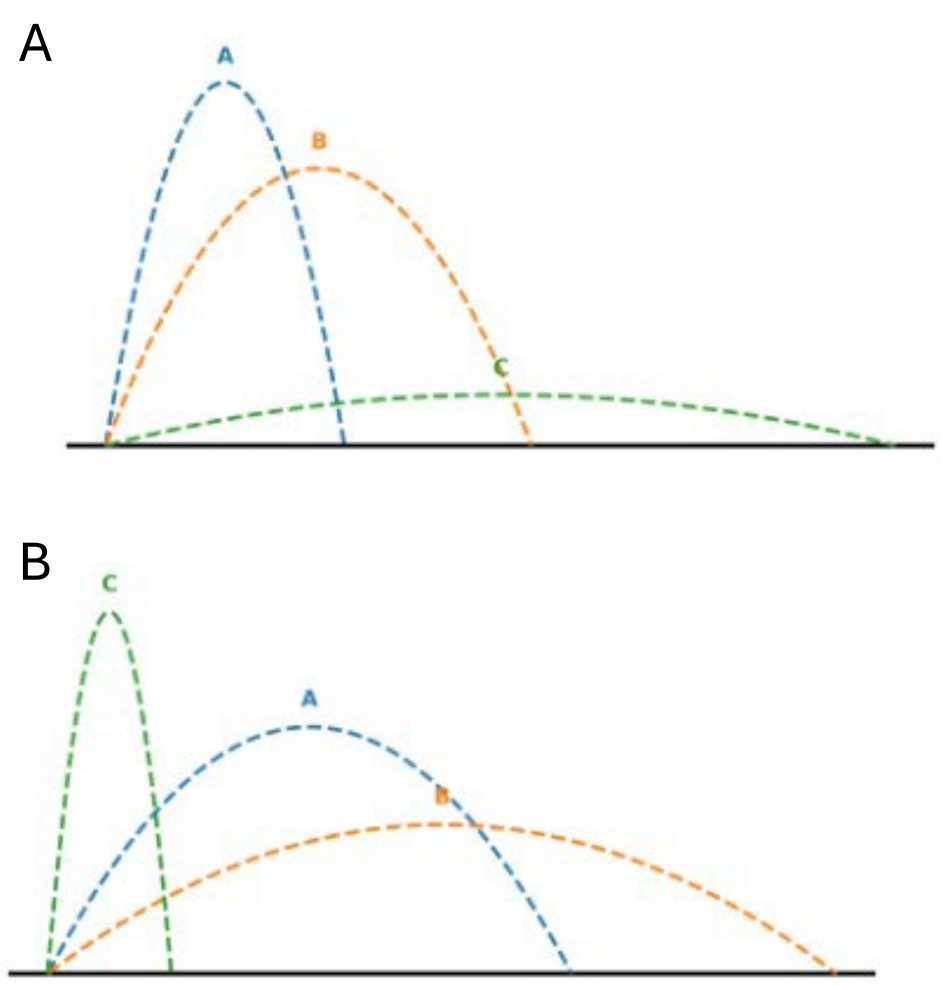

Figure 2: Diagrams for problem bank 2 created by chatGPT. A: Diagram for q-1. B: Diagram for q-9

### B. Comparison with single/simple prompt.

An example isomorphic problem generated by the single prompt approach for problem bank 1 is shown below:

A horse pulls a sledge across a snow-covered field using a harness that forms an angle of 40 degrees upward with respect to the horizontal. The sledge has a known mass, and the horse exerts a known pulling force. The sledge moves at constant speed. Find the coefficient of kinetic friction $\mu_k$. Round your answer to two significant figures.

The most significant shortfall of the single prompt output is that it completely neglected the instruction for generating numerical values for variables despite explicit instruction. All 10 versions generated had no numerical values associated with them.

An example problem generated by the simple prompt approach based on an example is shown here:

A robot drags a crate along an asphalt road at constant velocity.The force is **600 N**, applied at **20° upward**. The coefficient of kinetic friction is **0.45**.

Find the mass of the crate.

Answer: 140 kg

The main issue with this generation is that the correct answer should be 148.6 kg, ChatGPT simply hallucinated the answer without initiating python tool-use. Most of the problem answers were incorrect. The generated text is also noticeably shorter compared to the original example which was selected from problem bank 1. In fact, under the simple prompt condition chatGPT tends to write shorter text for each new problem generated.

For problem bank 2, we show in Figure 2 two example problem diagrams generated using the one-prompt approach. In both diagrams, one trajectory is underground. In diagram A, trajectory C is also invisible. This shows that under the one-prompt condition, the ability of the LLM to follow instruction is significantly compromised. In addition, the LLM somehow generated only 7 scenarios and 13 combinations, instead of the 10 scenarios and 26 combinations intended.

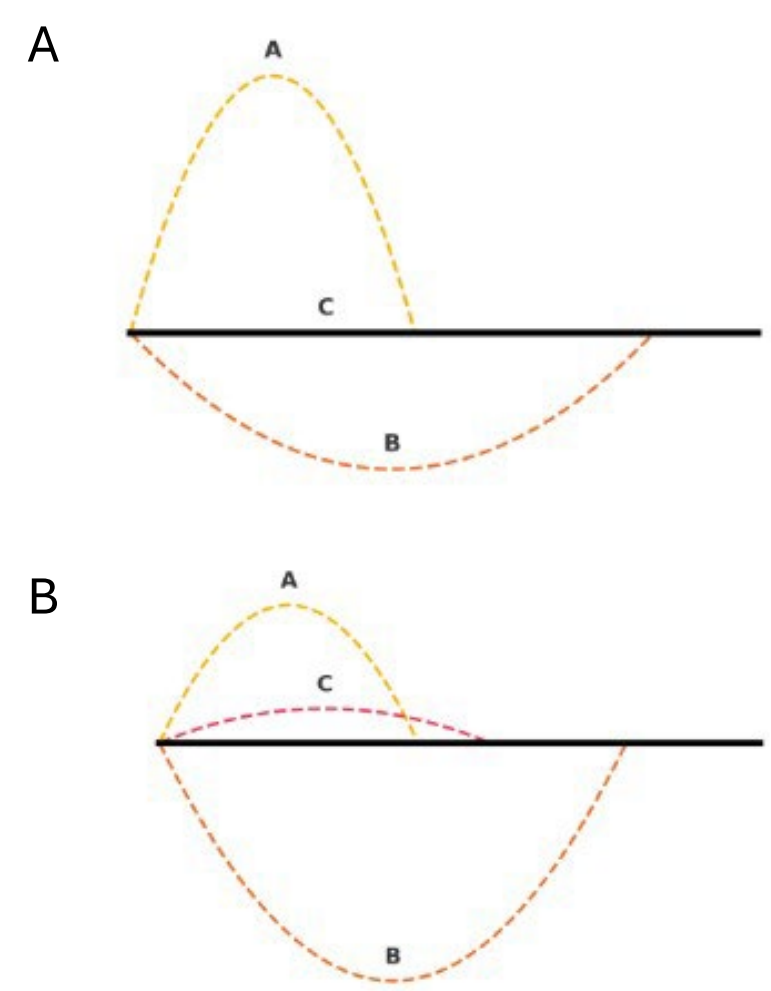

Figure 1: Two sample diagram generated by the simple prompt approach.

### C. Outcome Verification

Of the 20 total problems in Problem Bank 1, GenAI identified, and corrected, 6 problems that contain incorrect formula derivation in the solution text. The errors likely originated from the limited capabilities of GPT-4o at the time of the initial generation, as the second batch of 10 problems generated by Gemini Pro 2.5 with identical prompts did not contain any error.

For problem bank 2, GenAI identified three problems for which one of the distractors is equivalent to the correct answer. Human review of the problem bank found no additional errors. Both problem banks were also administered to students as both exam practice and exam items on a mid-term exam, and no further errors were reported by students.

## IV. CONCLUSIONS AND FUTURE DIRECTIONS

This study demonstrated using prompt-chaining and tool-use in ChatGPT to generate isomorphic problems with precisely controlled structural variations and diverse contextual elements. The resulting problems had high level of consistency in quality while having a wide variety of context, which is not possible to achieve by LLMs via single or simple prompts.

The prompt chaining approach offers the precision of traditional template-based AQG methods without the need for costly custom programs, while fully leveraging the generative capacity of LLMs. By incorporating the Python code interpreter, it also supports automatic generation of diagrams, addressing a common limitation of LLM-based item generation. Although developing a prompt-chain can require substantial effort up-front, the ability to reuse the same prompt chain for a near infinite number of isomorphic problems makes it highly efficient at scale. Existing prompt-chains can also be quickly modified to generate problem banks of similar problems.

In addition, the current approach also demonstrated the feasibility and efficacy of using GenAI to verify the quality of the generated items, which effectively removed erroneous items caused by hallucination of previous generation AI models.

As an exploratory effort, this work has several limitations that could be addressed in future research. First, the quality of generated problems was reviewed only by the author. Future studies should conduct more systematic quality review, and also collect student test data to evaluate psychometric properties of isomorphic problems. Second, we only focused on controlling structural variance, whereas future studies could explore more control on contextual variance such as using different styles of writing. Lastly, future studies should explore more complex diagram generation beyond the simple diagram in the current study.

Finally, the problem generation process could be further -automated by using generative AI to assist in the design of prompt-chain. If future studies could demonstrate consistent quality and reliability of this type of AQG strategy, it may even be possible to implement on-the-fly item generation for creating fully personalized assessments with infinite attempts.

## V. ACKNOWLEDGEMENTS

The current research is supported by NSF award EES 2421299 and the Gates Foundation.

## APPENDIX A: PROMPT CHAIN FOR GENERATION OF PROBLEM BANKS

### Problem Bank 1:

**prompt 1:**

I'm writing physics problems involving calculation of kinetic friction and balance of forces on a moving object. As a first step, please help me by doing the following:

Generate 10 different cases of an object being pulled or pushed by either a human, an animal, or a machine/vehicle across a surface. Indicate whether the pushing or pulling force must be angled upwards or downwards.

Here are some examples:

horse pulling a sledge on snow. upwards.
A person pushing a couch across carpeted floor. Downwards.
a person dragging a heavy luggage case. Upwards.
A towing truck dragging a damaged car on rough road. Upwards

**prompt 2:**

Now for each of those 10 cases, first generate the following set of values

1. Coefficient of kinetic friction, \mu_k. Between 0.1 and 0.9
2. Angle of the force \theta. Positive 10 degrees to 60 degrees if the force is upwards, negative 10 degrees to 60 degrees if the force is downwards.
3. Mass of the object being pulled or pushed, M. Appropriate for the object being pushed or pulled, in units of kg.
4. In addition, generate a generous estimation of the force that can be exerted by the human, animal or machine doing the pushing or pulling (not the object).

Next, calculate the magnitude of the force according to:

F = \mu * m *g/(cos(\theta) + \mu*sin(\theta)).

Check if the magnitude of F (in Newtons) is within the estimated range, and if the magnitude is positive. If yes, list the values. If not, re-generate the random numbers.

**prompt 3:**

For each of the 10 scenarios and the corresponding set of numbers, write a physics problem following those steps:

1. Choose either F, m, or \mu_k as the unknown variable.
2. Describe the situation and explain the known variables.
3. Imply that the object is moving/sliding at constant or uniform speed.
4. When stating the angle, explicitly state if it is upward or downward with respect to the horizontal, but not write any negative signs.
5. If the object is pulled by a rope or a chain or similar objects, state the angle as the rope/chain forms an angle of ….. with respect to the horizon
6. Ask students to calculate the unknown variable, and specify the unit. For example, find the force F in units of Newtons.
7. Specify significant figures required for the answer.

**prompt 4:**

Write concise student facing solutions for the first 5 problems. Each solution should include the following elements:

1. Explain that due to the object moving at constant velocity along a flat surface, the acceleration is zero in both directions, so the total force on both horizontal and vertical directions must both add up to zero according to Newton's second law of motion.

2. set positive x in the direction of the object's motion and positive y pointing up in the vertical direction.

3. Write down the sum of forces along the y-axis equal to zero expression, starting with \sum(\vec{F_y}) = ... = 0

4. Similarly, write down the sum of forces along the x axis equal to zero expression, using f to represent friction.

5. Find the friction force using the kinetic friction force equation, and find the normal force using the y-axis forces equation.

6. Find the unknown variable.

7. If the force is pointing downward, add a sentence clarifying that since the force is pointing below the x-axis, the numerical value of the variable \theta should be negative the value used in the problem.

Do not use numbers, instead, use languages similar to "First, notice that.... Next, we can define..., We can then write down...".

**prompt 5:**
Now please help me transform the first 5 problems into a YAML format, according to the example problem format below.

* The answer: value: field should contain the actual answer number from the data table above,
* The feedback: general field should contain the full solution generated above.
* Math equations and expressions should be written in LaTex and enclosed in `<latex></latex>` tags:

Note that prompt 5 was iteratively executed until all problems were transformed into a yaml format. The yaml format example is omitted here and can be found on

## Problem Bank 2:

**prompt 1:**
I would like to create different versions of a physics problem involving the trajectory of projectile motion of three objects, labeled A, B and C, that started from the same position from the ground and landed back on the ground. The question is about comparing the flight time of the three trajectories. First, please list all possible relations between the flight time of the three trajectories, for example, A > B > C, or A = B < C.
**prompt 2:**
Next, for each of the 13 flight time relation, I want to generate 2 possible relationships between the ranges of the three trajectories. The requirement is that:
1. The trajectories must be written in the same order as the flight time relation
2. For each relation in the flight time, replace with a different relation for Range. For example, replace ">" with either "<" or "=".
3. If a relation is in the non-determinant form such as "A < B > C", or "B > A < C" replace it with a different one according to the above two rules.
4. Avoid all equal relations such as "A = B = C" for the range.
Output the relations in a "long" table form with one column for flight time and one column for range

**prompt 3:**
Next step, for each row of the above table, come up with randomized parameters for the height of each trajectory, and the range of each trajectory. The requirements are: 1. the height parameter must be between 0.1 and 1, and the range parameter must be between 0.2 and 1.5 2. The relation between the height parameters should be the same as the flight time relations, and the relation between the range parameters should be the same as the range relations. 3. The variation between every two parameters should be greater than 0.2 and less than 0.8, unless they need to be equal to each other. Round all numbers to two decimal places after zero
**prompt 4:**
Now write a program using python and matplotlib to generate one diagram for each row in the above table, according to the parameters in the Table. In each diagram, the three trajectories will be represented by parabolic lines using dashed lines. All three trajectories should start from the same point on the x-axis and end on the x-axis (but not necessarily at the same location). Label the three trajectories with large and bold font, and use a thick line along the x-axis to represent "ground". Do not show axis or gridlines in the final diagram.
Name the diagrams q-1 through q-26, and create a downloadable zip file for all the diagrams
**prompt 5:**
Next, first add a column to the table named "Question", and label each row from q-1 to q-26, and retain only the flight time and range columns. Second, add two columns named "Distractor I" and "Distractor II". Each column should contain one random relation between A, B and C that is different from either the flight time or the range. Avoid non-determinant forms such as "A < B > C", or "B > A < C"
**prompt 6:**
Next, come up with 10 possible situations that involve projectile motion of three objects, all three projectiles should start from ground level and land on ground level. For example, three boys throwing rocks, three gunships firing guns, etc. Avoid objects that might be subjected to air resistance, such as arrows.
**prompt 7:**
All of them are good, randomly add the names of the 10 scenarios to the table above in a new column named "scenarios"
**prompt 8:**

Write the question body for each of the 10 scenarios above, stating that the trajectories are shown in the figure, and ask the students which of the following is the correct ranking of the time it takes for the objects to hit the ground. Do not write the choice items for now. Don't say "from longest to shortest".

**prompt 9:**

Now for each row in the table above, create the multiple choice problem with the choice items, according to the problem body above, and the four relations. The flight time relation is the correct answer. Write the multiple-choice problems according to the following yaml format example (the feedback section can be left blank for now):

## APPENDIX B: EXAMPLE OF COMPLETE PROBLEM BODIES AND SOLUTIONS

The problem bodies and solutions are listed in the yaml format used to store the generated problems, which contains latex code for mathematical equations.

### Problem Bank 1:

```
  - numerical:
      id: q1
      title: Worker Pushing a Wooden Crate
      points: 3
      text: |
        A worker is pushing a wooden crate across a concrete floor at a uniform speed.
        The coefficient of kinetic friction between the crate and the floor is <latex>0.36</latex>.
        The worker applies a force at an angle of <latex>12.74^\circ</latex> downward with respect to the horizontal.
        The required force exerted is <latex>465.64</latex> N. The acceleration due to gravity is <latex>9.81</latex> m/s².

        Find the mass of the crate in kilograms. Round your answers to two decimal places.

      answer:
          value: 117.67
          margin_type: percent
          tolerance: 3
      feedback:
          general: |
            First, notice that since the crate is moving at a constant velocity across a flat surface, its acceleration is zero in both the horizontal and vertical directions. According to Newton’s second law, the sum of forces in both directions must be zero.

            Next, we can define the positive <latex>x</latex>-direction as the direction of motion and the positive <latex>y</latex>-direction as pointing upward.

            We can then write down the force balance in the vertical direction:
            <latex>
            \sum \vec{F_y} = N - mg - F \sin(\theta) = 0
            </latex>
            where <latex>N</latex> is the normal force, <latex>m</latex> is the mass, <latex>g</latex> is the gravitational acceleration, and <latex>F</latex> is the applied force at an angle <latex>\theta</latex>.

            Similarly, for the horizontal direction:
            <latex>
            \sum \vec{F_x} = F \cos(\theta) - f = 0
            </latex>
            where <latex>f</latex> is the kinetic friction force.

            The friction force is given by:
            <latex>
            f = \mu_k N
            </latex>
            Substituting the expression for <latex>N</latex> from the vertical force equation:
            <latex>
            N = mg + F \sin(\theta)
            </latex>
            we get:
            <latex>
            f = \mu_k (mg + F \sin(\theta))
            </latex>
            Now, substituting <latex>f</latex> into the horizontal force equation:
```

```
      <latex>
      F \cos(\theta) = \mu_k (mg + F \sin(\theta))
      </latex>
      Solving for <latex>m</latex>:
      <latex>
      m = \frac{F \cos(\theta) - \mu_k F \sin(\theta)}{\mu_k g}
      </latex>
      The equations correctly account for the force being applied downward.

    on_correct: Good job!
    on_incorrect: Try again!
```

Problem Bank 2:

```
- multiple_choice:
    id: q-1
    title: Projectile Motion - Q-1
    points: 1
    text: 'Three engineering students test launchers that fire identical steel spheres
      from the same ground-level point. The three projectiles (A, B, and C) follow
      distinct parabolic paths and land back on the ground. Their trajectories are
      shown in the figure.

      Which of the following represents the correct ranking of their time of flight?'
    figure: Figures/q-1.png
    answers:
    - answer:
        text: C < A < B
        correct: false
    - answer:
        text: A > B > C
        correct: true
    - answer:
        text: A < B < C
        correct: false
    - answer:
        text: A = C < B
        correct: false
    feedback:
      general: 'For projectile motion, it can be shown that both the flight time and the height of the trajectory are directly proportional to the y-component of the initial velocity. So the higher the trajectory, the longer it takes for the projectile to hit the ground. The range, on the other hand, is determined by both the flight time and the x-component of the initial velocity, so one cannot compare the flight time directly from the range of the trajectory.'
      on_correct: ''
      on_incorrect: ''
```